\documentclass[ twocolumn,aps,prd,   
               preprintnumbers,numbers,sort&compress,nofootinbib,
                            showpacs,
               colorlinks,
               linkcolor=blue,
               citecolor=blue]{revtex4-1} 
   \newcommand{\exclude}[1]{}

\usepackage{graphicx,amsmath,amssymb,bm}
\usepackage{psfrag}
\usepackage{hyperref}

\begin{document}
\title{Isotropic Radio Background from Quark Nugget Dark Matter}

\author{Kyle Lawson and Ariel R. Zhitnitsky}
 \affiliation{Department of Physics \& Astronomy, University of British Columbia, Vancouver, B.C. V6T 1Z1, Canada}

\begin{abstract}
Recent measurements  by the \textsc{arcade2} experiment  unambiguously  show an excess in the isotropic radio background 
at frequencies below the GHz scale. We argue that this 
excess may be a natural  consequence of the interaction 
of visible and dark matter in the early universe if the dark matter consists of 
heavy nuggets of quark matter. Explanation of the observed radio band excess 
requires the introduction of no new parameters, rather we exploit   
the same dark matter model and identical normalization parameters 
to those previously used to explain other  excesses  of  
diffuse emission from the centre of our galaxy. These previously 
observed excesses include the
WMAP Haze of GHz radiation, keV X -ray emission and MeV gamma-ray radiation.
\end{abstract}
\maketitle

\section{Introduction}
The \textsc{arcade2} experiment has reported an excess in radio emission 
above the known \textsc{cmb} background at frequencies below the GHz scale 
\cite{Fixsen:2009xn}. When combined with earlier data 
\cite{Rogers:2008AJ}, \cite{Maeda:1999}, \cite{Haslam:1981}, \cite{Reich:1986}
the \textsc{arcade} result implies the existence of a radio background below the 
\textsc{cmb} peak which results in a measured antenna temperature that grows with 
a power index of approximately three \cite{Seiffert:2011apj}. This radio excess 
seems impossible to fit through reasonable modifications to the spectra of 
known background radio sources \cite{Ysard:2012pw}, \cite{Singal:2010}. 
Given the lack of a conventional astrophysical source  several attempts have 
been made to explain this excess through a contribution 
to the radio background from the annihilation of dark matter \cite{Fornengo:2011cn},
\cite{Fornengo:2011xk},\cite{Hooper:2012jc}, see also most recent analysis of the isotropic radio background in ref.\cite{Fornengo:2014mna}.

We argue that this radio excess 
naturally arises if the dark matter is composed of nuggets of quarks and antiquarks 
bound in a high density phase. The emission spectrum from these objects 
as a function of the visible matter density has been previously calculated and, 
if summed over the history of the universe, this radiation could easily 
provide the necessary isotropic radio band intensity. 
This is the main result of the present work.

We should emphasize that the model for dark matter to be used in our estimates below
has not been invented to explain the observed excess of radiation in radio bands. Rather, this model had been introduced with a completely different  motivation to be discussed below. 
After it's introduction the model was subsequently used to explain a number of other  excesses  of  diffuse emission from the centre of our galaxy  ranging over more than 10 orders of magnitude in energy scale. The  present work extends 
the application of this model from galactic to cosmological scales without changing  a single parameter in the model.

After a brief review of the dark matter model  in section \ref{model}    we analyze the cosmological  evolution of the dark matter nuggets 
in section \ref{thermodynamics}. We  estimate 
its potential contribution to the isotropic radio background in section \ref{spectrum} 
where we argue 
that such a contribution could account for the excess observed by the \textsc{arcade2} 
experiment below the GHz scale. 

\section{Dark Matter as Dense Quark Nuggets} \label{model}
It is known that dark matter dominates over the visible matter of the 
universe, with which it interacts only weakly, by a factor of approximately 
five. While the behaviour of dark matter on the scale of galaxies and above is 
well understood there is no generally accepted model for its microscopic 
nature. The majority of dark matter models assume the existence of a 
new fundamental particle whose properties are to be inferred from the 
behaviour of the dark matter itself. This paper will consider an alternative 
model in which the dark matter is composed of nuggets of standard model 
quarks and antiquarks bound in a high density phase. This class of models 
(which includes for example Witten's stranglets \cite{Witten:1984rs}) is allowed 
due to the fact that all dark matter detection experiments are sensitive 
not to the dark matter cross section but to the product of the cross section with 
the number density. Since only the dark matter mass density, both locally and globally, 
is experimentally measured this amounts to a limit on the dark matter 
cross section per mass $(\sigma/M)$. Consequently, composite strongly interacting  objects  
may be able to constitute the dark matter, despite 
having a significant interaction cross section, provided their mass 
is sufficiently large. 

\subsection{The Model} \label{model1}
In this work we will consider a dark matter model presented in 
\cite{Zhitnitsky:2002qa,Oaknin:2003uv} and reviewed in \cite{Lawson:2013bya}. The basic idea  of this model
is that nuggets of dense matter and
antimatter  form at the same \textsc{QCD} phase transition as
conventional baryons (neutrons and protons), providing a natural
explanation for the similar scales $\Omega_{\textsc{DM}} \approx
5\Omega_{\textsc{B}}$.  Baryogenesis proceeds through a charge
separation mechanism: both matter and antimatter nuggets form, but the
natural \textsc{CP} violation due to the so-called QCD $\theta$ term  (which was
of order unity $|\theta|\sim 1$ during the QCD  phase
transition)  drives the formation of more antimatter nuggets than
matter nuggets, resulting in the leftover baryonic matter that forms the
visible matter today. In different words, while the total baryon charge 
of the universe is zero in this model, different components are not equally 
distributed between nuggets, antinuggets and the visible matter, i.e.  
$B_{\text{universe}} =  B_{\text{nugget}}
    + B_{\text{visible}}-\bar{B}_{\text{antinugget}}=0$.
It is crucial for this mechanism that \textsc{CP} violation can drive
charge separation in the early universe during the QCD phase transition 
while today the same source of \textsc{CP} violation can be safely neglected. 
This is a result of the well known resolution of the so-called strong \textsc{CP} 
problem with the dynamical axion in which $\theta\sim 1$ during the QCD  phase
transition becomes $\theta\simeq0$ at a later epoch, 
see the original papers \cite{Zhitnitsky:2002qa,Oaknin:2003uv} and review \cite{Lawson:2013bya} for relevant 
references and details.  
Note, that  the  idea of a charge separation mechanism due to the local violation of  
\textsc{CP} invariance can be experimentally tested  at the Relativistic Heavy Ion Collider (RHIC) and the LHC. We include a few comments  and relevant references, including some references to recent experimental results supporting the basic idea,  in section \ref{conclusion}.   
 
If this proposal is to explain the both the 
absence of primordial antimatter in the hadronic phase and the 
observed visible to dark matter ratio 
then the matter composition of the universe should be roughly one part visible 
matter, two parts dark matter in the form of quark nuggets and three parts 
dark matter in the form of antiquark nuggets. This composition would explain 
the dominance of visible matter over antimatter and the approximately 1:5 visible 
matter to dark matter ratio, i.e. $\bar{B}_{\text{antinugget}}$:$B_{\text{nugget}}$:$B_{\text{visible}}\simeq
$ 3:2:1.  Though this ratio cannot be presently computed exactly, a simple   
estimate  suggests that the three components in this ratio should be order 
of one  as a result of strong dynamics during the QCD phase transition when 
all parameters, including $\theta$ are order of one. 

Unlike conventional dark matter candidates, dark-matter/antimatter
nuggets are strongly interacting but macroscopically large.  
They do not contradict any of the many known observational
constraints on dark matter or
antimatter  for three reasons: 1) They carry a huge (anti)baryon charge 
$|B| \sim10^{24}$, and so have an extremely tiny number
density; 2) The nuggets have nuclear densities, so their effective interaction
is small $\sigma/M \sim 10^{-10}$ ~cm$^2$/g,  well below the typical astrophysical
and cosmological limits which are on the order of 
$\sigma/M<1$~cm$^2$/g; 3) They have a large binding energy 
such that baryon charge  in the
nuggets is not available to participate in big bang nucleosynthesis
(\textsc{bbn}) at $T \approx 1$~MeV. Weakness of the visible-dark matter 
interaction is achieved 
in this model due to the small parameter $\sigma/M \sim B^{-1/3}$ 
rather than due to a weak coupling 
of a new fundamental field with standard model particles. 

While the observable consequences of this model are on average strongly suppressed  
by the low number density of the quark nuggets the interaction of these objects 
with the visible matter of the galaxy will necessarily produce observable 
effects. The interaction between the nuggets and their environment is governed 
by well know nuclear physics and basic \textsc{qed}. As such their 
observable properties contain relatively few tunable parameters allowing several 
strong tests of the model to be made based on galactic observations. It is found that 
the presence of quark nugget dark matter is not only allowed by present 
observations but that the overall fit to the diffuse galactic emission  spectrum 
across many orders of magnitude in energy may be improved by their inclusion. 
This includes a number of frequency bands where some excess  of emission,  
which cannot be explained by conventional astrophysical sources have been observed. 
Here we briefly review the status of several galactic observations which 
may support the idea of quark nugget dark matter. 

\subsection{Excess emissions from the galaxy as viewed from the DM model}
\label{sec:gal_emis}
The model makes unambiguous
predictions about  the processes ranging over more than 10 orders of
magnitude in scale.  The basic picture involves the antimatter
nuggets -- compact cores of nuclear or strange quark matter surrounded
by a positron cloud, the  ``electrosphere"  with a profile which can be computed using conventional and well-established physics.
Incident matter will annihilate on
these nuggets producing radiation at a rate proportional to the
annihilation rate, thus scaling as the product $\rho_{V}(r)\rho_{DM}(r)$ of
the local visible and dark matter densities.  This will be greatest in
the core of the galaxy.  To date, we have considered several independent
observations of diffuse radiation from the core of our galaxy:

${\bullet}$  The annihilation of the positrons of the  ``electrosphere" with the electrons 
of the interstellar medium will result in formation of a positronium which consequently decays. The decay will produce a 511 keV  line and the 
associated three photon continuum.  Such a spectral feature is in fact observed 
and has been studied by the \textsc{SPI Integral} observatory \cite{Jean:2005af}. 
In our model the properties of the 511 line are naturally 
explained: The strong peak at the
galactic centre and extension into the disk must arise because the
intensity follows the distribution $\rho_{V}(r)\rho_{DM}(r)$ of
visible and dark matter densities~\cite{Oaknin:2004mn}, 
\cite{Zhitnitsky:2006tu}. This should be contrasted with the $\sim \rho_{DM}^2(r)$ 
scaling which arises in the case of dark matter annihilation or $\sim \rho_{DM}(r)$ 
in the case of decaying dark matter. A distribution $\sim \rho_{V}(r)\rho_{DM}(r)$ 
obviously implies that the predicted emission will be asymmetric,
extending  into the disk from the galactic centre as it tracks the visible matter. 
Apparently, such an asymmetry has been observed ~\cite{2008Natur.451..159W}, 
see also the review paper \cite{Prantzos:2010wi}.
The intensity of the emission of the 511~keV line was used   to normalize the line 
of sight integral $ \int dr\rho_{V}(r)\rho_{DM}(r)$  in our analysis of all 
other galactic emissions.
 
${\bullet}$ Positrons closer to the quark matter surface can carry energies up 
to the nuclear scale. Some  galactic electrons are able to penetrate to a sufficiently 
large depth that they no longer produce  the characteristic positronium decay 
spectrum but a direct $e^-e^+ \rightarrow 2\gamma$ emission spectrum 
\cite{Lawson:2007kp}. The transition between these 
two regimes is determined by conventional physics and allows us to  compute   
the strength and spectrum of the MeV scale emissions relative to 
that of the 511~keV line \cite{Forbes:2009wg}. Observations 
by the \textsc{Comptel} satellite show an excess above the galactic 
background \cite{Strong:2004de}  consistent with our estimates. 

${\bullet}$ Galactic protons may also annihilate with the antimatter 
comprising the quark nuggets. The annihilation of a proton within a 
quark nugget will produce hadronic jets which cascade down into 
lighter modes of the quark matter. If the energy in one of these jets reaches the 
quark surface directly it will excite the most weakly bound positron states near the 
surface. These excited positrons rapidly loose energy to the strong electric field 
near the quark surface. This process results in the emission of Bremsstrahlung 
photons at X-ray energies \cite{Forbes:2006ba}. Observations by the 
\textsc{Chandra} observatory indeed indicate an excess in X-ray emissions from 
the galactic centre  \cite{Muno:2004bs} with the intensity and spectrum 
consistent with our estimates \cite{Forbes:2006ba}.

${\bullet}$ The annihilation of visible matter within the nuggets heats them above the 
background temperature. The thermal spectrum from the nuggets may be 
predicted based on the emission properties of the electrosphere along with the 
annihilation rate at various positions within the galaxy \cite{Forbes:2008uf}.
The majority of this thermal energy is emitted at the eV scale where it 
is very difficult to observe against the galactic background. However the 
emission spectrum will extend down to the microwave scale where it may 
 be responsible for the ``\textsc{wmap} haze" \cite{Finkbeiner:2003im}.  

\subsection{Prospects for ground based detection}
While the galactic backgrounds discussed above may offer indirect support 
for our proposal the possibility of direct detection through earth based observations 
should also be considered. The corresponding questions have been discussed recently in refs \cite{Lawson:2010uz,Lawson:2012vk,Gorham:2012hy}.

The scale of any observable 
consequences will be determined by the flux of nuggets through the earth. The 
local dark matter density is roughly $1GeV/cm^3$ and has a velocity around the 
galactic scale of $v \sim 200 km/s$. Taking these values the mean baryonic 
charge of the nuggets then sets the flux expected at the earth's surface.  
\begin{equation}
\label{eq:flux}
\frac{dN}{dA ~ dt} = nv \approx \left( \frac{10^{25}}{B} \right) km^{-2} yr^{-1}
\end{equation}
Thus, if the baryonic charge is on the order of $10^{25}$ we should anticipate a 
flux of one nugget per square kilometer per year with the flux dropping for 
larger (and thus less abundant) nuggets. This flux is well below the sensitivity 
of any conventional dark matter searches but, if the energy deposited by such an 
event is sufficiently large there may still be observational consequences. 

A nugget of antimatter crossing the atmosphere will annihilate any atmospheric  
material which lies in its path. The atmospheric mass involved is substantially 
smaller than that of the nugget and thus the nugget looses virtually none of its 
original momentum and charge, and the total energy released is simply proportional to 
the amount of atmospheric material swept up. We may  thus 
estimate the energy released in such an event as, 
\begin{equation}
\label{eq:tot_energy}
E_{tot} = 2X_{at}c^2 \pi R_n^2 \approx 10^{7}J \left( \frac{R_n}{10^{-5}cm} \right)^2.
\end{equation} 
where $X_{at} \sim 1kg/cm^3$ is the atmospheric depth. The majority of this 
energy will be thermalized inside the nugget however some may be deposited 
in the atmosphere in an observable form. While the energy scale is similar 
to that released in a typical meteor event the majority of this energy will be released 
quiet low in the atmosphere limiting direct visual observation. It has been  also argued \cite{Lawson:2010uz}  that only a fraction of all molecules incident on the nugget actually annihilate. Thus, the expression (\ref{eq:tot_energy}) represents a maximum energy available to the shower with the actual value likely to be several orders of magnitude smaller.

Recent work has considered the possibility that large scale 
cosmic ray detectors 
may be capable of observing quark nuggets passing through the earth's
atmosphere either through the extensive air shower such an event 
would trigger \cite{Lawson:2010uz} or through the geosynchrotron 
emission generated by the large number of secondary particles
\cite{Lawson:2012vk}. It has also been suggested that the \textsc{anita} 
experiment \cite{Gorham:2008dv} may be sensitive to the radio band 
thermal emission generated by these objects as they pass through the 
antarctic ice \cite{Gorham:2012hy}. These experiments may thus be 
capable of adding direct detection capability to the indirect evidence 
discussed above. 

On entering the earth's crust the nugget will continue to deposit energy along 
its path, however this energy is dissipated in the  surrounding rock and is unlikely 
to be directly observable. Generally the nuggets carry sufficient momentum to travel 
directly through the earth and emerge from the opposite side however a small fraction 
may be captured and deposit all their energy. In \cite{Gorham:2008dv} the 
possible contribution of energy deposited by quark nuggets to the earth's 
thermal budget was estimated and found to be consistent with observations. 

The nuggets' thermal spectrum which may be relevant to direct detection searches 
arises as the annihilation of atmospheric or ice molecules colliding with the nugget 
raise its temperature causing it to emit in the radio band. While the 
density is considerably lower, the same process will result in radio 
emission from astrophysical regions of relatively high density. A
critical aspect of the analysis presented here, as well as that 
conducted in \cite{Forbes:2008uf} and \cite{Gorham:2012hy}, is the 
relative flatness of the nugget emission spectrum with respect to 
black body radiation.  To discuss this further we now turn 
to a more quantitative description of the thermal spectrum generated 
by a nugget of quark matter.

\section{Nugget Thermodynamics}\label{thermodynamics}
As discussed above, thermal emission from the nuggets may provide a 
component of the \textsc{wmap} haze. In investigating this proposal 
the thermal emission spectrum of a nugget of quark matter was worked out in 
\cite{Forbes:2008uf} where the spectrum was found to be, 
\begin{eqnarray}
\label{eq:the_spec}
\frac{dE}{dt~dA~d\nu} &=&\\
  \frac{4}{45}\cdot \frac{T_N^3 \alpha^{5/2}}{\pi}&\cdot&
\sqrt[4]{\frac{T_N}{m_e}} \left(1 + \frac{h\nu}{T_N} \right) e^{-h\nu / T_N} 
f\left( \frac{h\nu}{T_N} \right) \nonumber
\end{eqnarray} 
where $T_N$ is the temperature of the quark nuggets and the function 
$f(x)$ is defined to be, 
\begin{eqnarray}
f(x) &\equiv& 17 - 12\ln(x/2) ~~~x<1\nonumber \\
&\equiv&  17 + 12 \ln(2) ~~~x>1
\end{eqnarray} 
It should be noted that the leading order dependence of this spectrum on 
frequency is relatively weak resulting in a near flat spectrum at energies 
below the nugget temperature. 

The derivation of expression (\ref{eq:the_spec})  did not account 
for the finite size of the nuggets. When finite-size effects are considered 
they impose a low energy cutoff on the thermal spectrum. 
The spectrum was derived based on the scattering of positron states 
of arbitrary long wavelength, in reality these states must be limited to 
length scales below the size of the nugget. These considerations  
significantly alter the relevant physics on momentum exchange 
scales of order $\delta p \sim \hbar/R$, below which positron 
scattering will be suppressed. Taking a typical nugget radius of 
$R\sim10^{-5}$cm implies a cutoff in the photon spectrum at energies 
$E_{cut} \sim \delta p^2 /2m \sim 10^{-4}$eV. As the photons generated in 
positron scattering need not remain bound to the nugget a similar 
wavelength restriction is not required for the photon states. This implies 
that the resulting low energy photons will not be in strict thermal equilibrium 
with the electrosphere. This emission cutoff was mentioned 
in \cite{Forbes:2008uf}, but was below the energy range considered in that 
analysis. As we are here interested in the nugget contribution to the radio 
background at frequencies below this scale the low energy cutoff 
must be taken into account. The 
exact form of this cutoff will depend on the dynamics of the electrosphere 
in a more complicated way than is appropriate for the present analysis. 
For this reason we will estimate the effect by simply imposing 
 a hard cutoff at the energy scale $E_{cut}$ estimated above. 

The basic proposal presented here is that 
while the nuggets make a contribution to the isotropic background which is 
much smaller than that of the \textsc{cmb} at its peak ($\sim 160~$Ghz) the 
thermal spectrum of the \textsc{cmb} falls as $\sim \nu^{3}$ below peak while 
the nugget contribution remains essentially constant coming to dominate 
at low energies and resulting in an observed isotropic background temperature that 
seems to grow with the third power of frequency. 

To expand on this basic idea we must estimate the temperature evolution of 
the nuggets over the history of the universe. To do this we compare the rate at 
which thermal energy is emitted with the rate at which visible matter deposits 
energy in the nugget through annihilations. 
The integrated thermal radiation from a nugget is given by, 
\begin{equation}
\label{eq:tot_flux}
\frac{dE}{dt} = 4\pi R_N^2 \frac{16}{3\pi} \alpha^{5/2}T_N^4\sqrt[4]{\frac{T_N}{m_e}}
\end{equation}
This rate should be proportional to the flux of visible matter onto the quark nugget. 
\begin{equation}
\label{eq:matter_flux}
\Phi_{vis} = \rho_{vis}c^2 v \pi R_N^2
\end{equation}
The velocity of visible matter drops proportional to the scale factor while the 
visible matter density falls as $a^{-3}$. 
\begin{equation}
\Phi_{vis}(z) =  \rho_{0}c^2 v_0 \pi R_N^2 (1+z)^4 
\end{equation}
As only a fraction of the energy released through annihilations is  
actually thermalized in the nugget expressions 
(\ref{eq:tot_flux}) and (\ref{eq:matter_flux}) are equal up to some multiplicative 
constant less than one. While other mechanisms for energy loss 
exist (such as those leading to the various galactic sources described in 
({\ref{sec:gal_emis})) thermal emission is expected to be the 
dominant cooling mechanism. This is due to the fact that the 
approximately $2~ $GeV released by the annihilation of a proton 
within the quark matter rapidly cascades down to many low energy 
excitations of the light modes of the quark matter. By the 
time this energy is transported to the quark surface the 
individual excitations do not have sufficient energy to overcome 
the nuclear scale binding energy of the quark matter and produce 
significant levels of particle emission. The same strong binding force  prevents the
anti-baryon charge to escape the system. 

The numerical coefficient relating the annihilation rate and 
thermal emission is not 
a free parameter and it can,  in principle,  be computed,  the details of such a 
computation will be commented upon below.  
This procedure results in a thermal evolution of the form 
\begin{equation}
\label{eq:T_evol}
T(z) = T_{LS} \left( \frac{1 + z}{1 + z_{LS}} \right)^{17/4}.
\end{equation}
Here the parameter $T_{LS}$ is the effective radiating temperature of the nuggets 
(as it appears in expression (\ref{eq:the_spec})) at the time of last 
scattering (i.e. during \textsc{cmb} formation). This temperature is 
calculable in principle, but is dependent on 
the fraction of visible matter which is converted to thermal energy in the 
nuggets after its annihilation. Rather than attempting a direct calculation of this temperature 
we shall simply treat it as a phenomenological parameter to be 
fit by the data. It should be noted that a very similar calculation was 
carried out   in \cite{Forbes:2008uf} where it was estimated that, to provide 
the observed \textsc{wmap} galactic haze  the nuggets must have a 
temperature at the eV scale in the galactic centre. As we estimate below, the 
present density of the visible matter at the centre of galaxy and the corresponding 
density at the time of the last
scattering are about the same. Consequently, $T_{LS}$ is expected to be 
at  the same eV scale estimated  in \cite{Forbes:2008uf}.  As mentioned above, we opted to treat $T_{LS}$ as a fitting parameter which we expect to be in eV scale range.
\exclude{If we assume that the 
entire haze is due to the nuggets then consistency with the \textsc{wmap} data
would require a nugget temperature at last scattering of $T_{LS} \sim 0.5$eV. 
We will take this as a rough estimate in what follows. }

To get a basic idea of the scale at which we may expect the isotropic 
radio background we may compare it to the foreground galactic emission 
which we claim is visible as the \textsc{wmap} haze. In the environment of 
the galactic centre with the nuggets carrying a typical galactic velocity 
the matter flux onto the nugget scales as 
\begin{equation} 
\label{galaxy}
\rho_{vis}v \approx 300 \frac{\rm GeV}{\rm cm^3}\cdot 2\cdot 10^{7}\frac{\rm cm}{\rm s} 
= 6\cdot  10^9 \frac{\rm GeV}{\rm cm^2 s}, 
\end{equation}
where we use the average numerical values adopted in \cite{Forbes:2008uf}.
A similar calculation at the time of last scattering gives 
\begin{equation}
\label{ls}
\rho_{vis} v = \rho_c \Omega_{vis} \sqrt{\frac{2T}{m_P}}c (1+z)^3 
\approx 3\cdot10^{8}  \frac{\rm GeV}{\rm cm^2 s}. 
\end{equation} 
The similarity of these  scales  (\ref{galaxy}) and (\ref{ls}), and the 
relatively weak dependence of temperature on the matter flux, indicate
that the initial intensity of this 
cosmological background should have been similar to that of the present day 
galactic spectrum.  As the universe expands and cools the nugget's 
contribution to the isotropic background falls off quickly and photons 
produced at early times redshift to lower energies. This process means 
that, at the present day \textsc{cmb} peak the nugget background contribution 
to the spectrum is vanishingly small. However, at frequencies below the 
\textsc{cmb} peak the flatness of the nugget spectrum causes the background 
to eventually emerge as the dominant isotropic radio source.  

\section{Emission Spectrum}\label{spectrum}
Having established the temperature evolution of the nuggets as well as their 
thermal spectrum we may now determine the present day radio background 
due to the presence of quark matter nuggets over the history of the universe. 
This is accomplished by integrating the source density over all redshifts 
correcting for the redshifting of photon frequency after emission.
\begin{equation} 
\label{eq:intens_int}
I(\nu) = \frac{c}{H_0} \int \frac{dz}{(1+z)h(z)}\frac{\rho_{DM}}{M_N} 
\frac{dE}{d\nu ~ dt} \left[ \nu (1+z), T(z) \right]
\end{equation}
For simplicity this analysis we will ignore the reheating of matter during structure 
formation at late times. Obviously a full treatment would involve the clustering 
of dark matter into the structure observed in the universe today as was done 
in \cite{Hooper:2012jc} and \cite{Fornengo:2011cn}. However these considerations 
go beyond the scope of this paper which seeks only to demonstrate the feasibility 
of this mechanism of producing an isotropic radio background. In this simplification 
we are aided by the fact that the density of visible matter must be quite strongly 
enhanced before the nuggets begin to again contribute to the radio background. By this time 
the visible matter in the galaxies will also be generating radiation at these bands 
and much of the radio signal will be lost as a ``haze" contribution to the host galaxies.    

We may estimate the numerical impact of late time reheating by considering  
modifications of the thermal dependence given in (\ref{eq:T_evol}). For example 
suppose that at redshift less than z=10 we allow one tenth of the dark matter 
to reheat to a temperature of 1eV (the present day value in the galactic centre.) 
This will give produce two terms in the integrand of (\ref{eq:intens_int}) one 
with a temperature falling as before and the other fixed at 1eV below z=10. 
The contribution of this second term does not strongly impact the value of the 
total integral which is heavily weighted towards early times. Numerically this 
procedure is found to introduces variations in the predicted 
antenna temperature on the 
order of a few percent across the frequency range considered here. 
Such a variation is considerably smaller than the uncertainty in the initial 
nugget temperature $T_{LS}$. Including structure formation would of 
course produce a more precise result, and would allow for the consideration 
of small anisotropies in the resulting radio background but the present estimate 
will be taken as sufficient for the purposes of this work.
 
With this simplifying  assumption the integral (\ref{eq:intens_int}) 
may be performed using the nugget 
temperature dependence given in (\ref{eq:T_evol}). The results of 
this integration are plotted in figure \ref{fig:Rad_spec} for a 
specific choice of the mean nugget baryon charge of $B\simeq 10^{25}$, 
the initial temperature of the nuggets was then estimated by a fit 
to the \textsc{Arcade2} data. Similar fits are possible across the 
range of allowed nugget sizes, with larger nuggets requiring a 
higher initial temperature to fit the data. In all models considered the 
best fit temperature fell in the $T \lesssim 1$eV. 
\begin{figure}[t]
\begin{center}
\includegraphics[width = 0.5\textwidth]{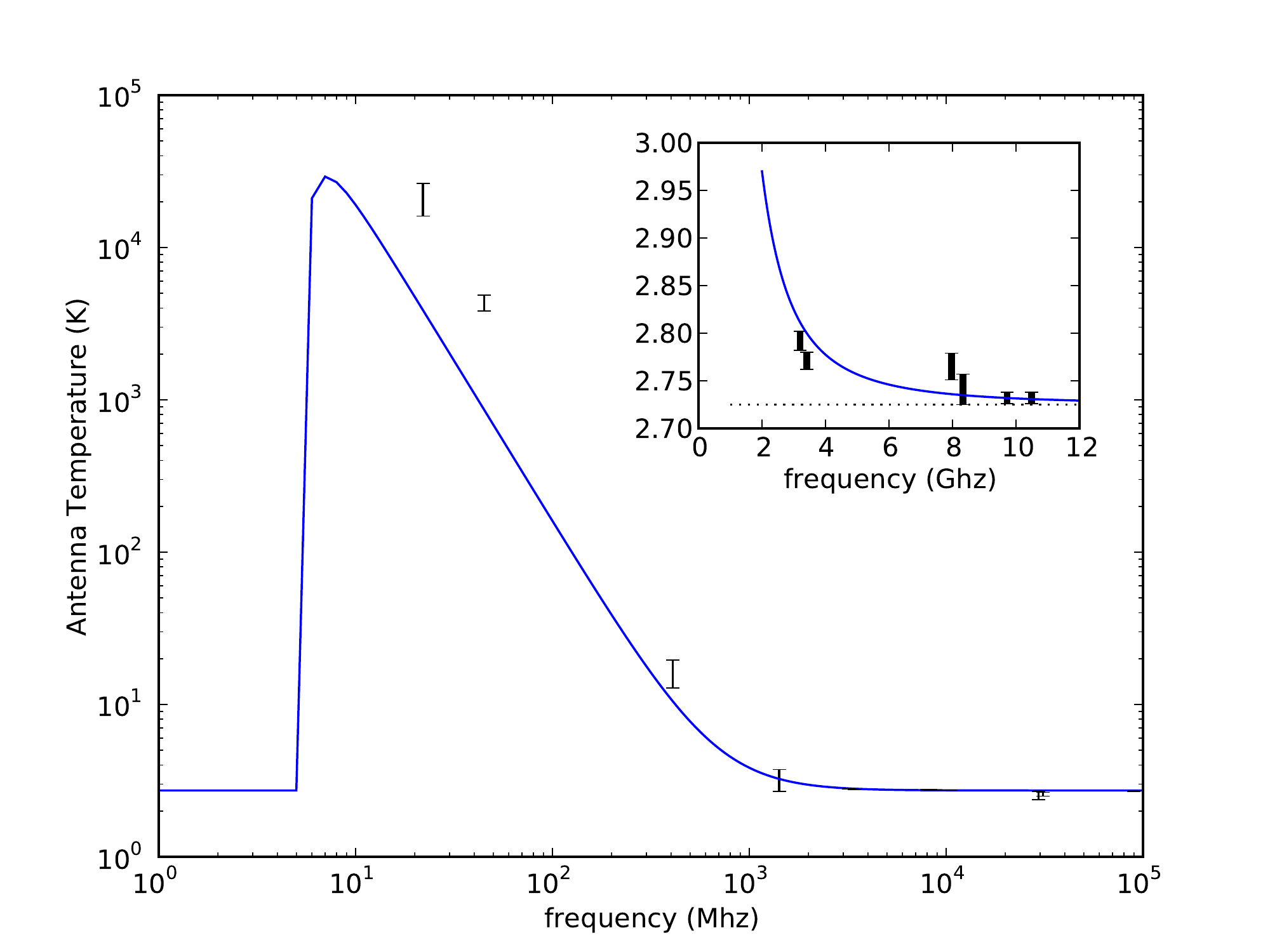}
\caption{Predicted antenna temperature assuming a mean quark nugget 
baryon number $B\sim 10^{25}$ and a
nugget  temperature of 0.2eV at the time of \textsc{cmb} 
formation. Also plotted are the data points from the radio band observations 
cited in the text. The insert is a close up on the \textsc{Arcade2} 
data points for which an excess, above the \textsc{CMB}, is observed.}
\label{fig:Rad_spec}
\end{center} 
\end{figure}    
From figure \ref{fig:Rad_spec} it can be seen that the nuggets make little 
contribution at frequencies near the CMB peak (associated with photons 
emitted at $\sim$ 0.3eV at the time of last scattering) but, due to the relative 
flatness of the  nugget emission spectrum (\ref{eq:the_spec}), come to 
dominate below Ghz frequencies where the thermal spectrum of the CMB 
photons fall as $\sim \nu^3$. This behaviour naturally explains the spectral 
index of $\gamma \simeq 3$ described in \cite{Seiffert:2011apj}. 

\section{Conclusion}\label{conclusion}
The main result of this work can be formulated as follows. 
If the dark matter consists of heavy nuggets of quark 
matter in a high density phase then these objects may have a more complex 
thermal history than dark matter consisting of a new 
fundamental particle. We have estimated the thermal evolution of quark matter nuggets  
from the time the universe first becomes transparent and 
determined the impact of this additional source of diffuse radio emission 
on the present day background. This analysis finds that at energies near the
\textsc{cmb} peak the nugget contribution to the radio background is several 
orders of magnitude below that of the thermal \textsc{cmb} spectrum. However
the \textsc{cmb} spectrum falls of at frequencies below peak much faster than 
that of the nuggets such that, at frequencies below roughly a Ghz, they come to 
give the dominant contribution to the isotropic radio background. 
As such the presence of dark 
matter in the form of quark nuggets offers a potential explanation of the 
radio excess observed by \textsc{arcade2}.

We emphasize that the only fitting parameter, $T_{LS}$ entering 
the final result presented in Fig.\ref{fig:Rad_spec} is not really a free parameter, 
and in principle can be computed 
as it is determined by the conventional well-established physics. We presented 
the order of magnitude estimate $T_{LS}\sim 1$ eV, and the fit
plotted  on Fig.\ref{fig:Rad_spec} is consistent with this estimate.  One should add that this  dark matter proposal   may explain  a number of  ``apparently unrelated" puzzles. 
All these puzzles  strongly suggest (independently)  the presence of some source of
excess diffuse radiation in different bands ranging over 13 orders of magnitude in 
frequency. The new element which we advocate in this paper is that the 
same dark matter model which offers a source for these previously 
discussed excesses of diffuse emission can also explain that  
observed by \textsc{arcade2} in radio bands. In this case the emission originates 
primarily from very early times with $z\sim 10^3$ in contrast with our previous 
applications which have analyzed only present day galactic emissions.  

 Finally, what is perhaps more remarkable is the fact that the key assumption of this dark matter model,  the charge separation effect reviewed in section \ref{model1},  
can be experimentally tested in heavy ion collisions, where a similar 
$\cal{CP}$ odd environment with $\theta\sim 1$ can be achieved, see  
section IV in ref.\cite{Kharzeev:2007tn} for the details. 
In particular, the local  violation  of the $\cal{CP}$ invariance  observed at RHIC (Relativistic Heavy Ion Collider)\cite{Abelev:2009tx} and LHC (Large Hadron Collider)\cite{Abelev:2012pa}  have been interpreted in  \cite{Kharzeev:2007tn,Zhitnitsky:2010zx,Zhitnitsky:2012im} as an outcome of  a charge separation mechanism in the presence of the induced $\theta\sim 1$   resulting from  a collision. The difference is of course that
 $\cal{CP}$ odd  term with  $\theta\sim 1$  discussed in cosmology describes a theory  on the horizon scale, while $\theta\sim 1$ in heavy ion collisions is correlated on a size of the colliding nuclei. 
 
\section*{Acknowledgements}
We are thankful to Roberto Lineros, Nicolao Fornengo, Marco Regis and Marco Taoso for correspondence on 
questions related to the clustering and structure formation.  
This research was supported in part by the Natural Sciences and Engineering Research
Council of Canada. KL is supported in part by the UBC Doctoral Fellowship program.


\end{document}